\documentclass[conference,10pt]{IEEEtran}

\usepackage[english]{babel}
\usepackage[T1]{fontenc}
\usepackage{cite,url,color} 
\usepackage{graphics,amsfonts}
\usepackage{epstopdf}
\usepackage[pdftex]{graphicx}
\usepackage[cmex10]{amsmath}
\usepackage[utf8]{inputenc}
\usepackage{enumitem}

\usepackage{tikz}
\usetikzlibrary{automata,positioning,chains,shapes,arrows,patterns}
\usepackage{pgfplots}
\usetikzlibrary{plotmarks}
\newlength\fheight
\newlength\fwidth
\pgfplotsset{compat=newest}
\pgfplotsset{plot coordinates/math parser=false}

\usepackage{array}
\usepackage[font=footnotesize]{caption}
\usepackage[font=footnotesize]{subcaption}

\usepackage[top=1.5cm, bottom=2cm, right=1.6cm,left=1.6cm]{geometry}
\usepackage{indentfirst}

\usepackage{times}

\usepackage{breqn}
\usepackage{booktabs}
\usepackage{threeparttable}
\usepackage{tabularx}
   \usepackage{multirow}
   \usepackage{booktabs}

   \usepackage{array, blindtext}
   \usepackage{wrapfig}
\usepackage{pdfpages}

\usepackage{etoolbox}
\usepackage{etex}

\usepackage[acronyms,nonumberlist,nopostdot,nomain,nogroupskip]{glossaries}
\usepackage{glossary-mcols}
\newacronym{3gpp}{3GPP}{3rd Generation Partnership Project}
\newacronym{5g}{5G}{5th generations}
\newacronym{aimd}{AIMD}{Additive Increase Multiplicative Decrease}
\newacronym{am}{AM}{Acknowledged Mode}
\newacronym{amc}{AMC}{Adaptive Modulation and Coding}
\newacronym{aqm}{AQM}{Active Queue Management}
\newacronym{awgn}{AGWN}{Additive White Gaussian Noise}
\newacronym{balia}{BALIA}{Balanced Link Adaptation}
\newacronym{bdp}{BDP}{Bandwidth-Delay Product}
\newacronym{bf}{BF}{Beamforming}
\newacronym{cc}{CC}{Congestion Control}
\newacronym{cdf}{CDF}{Cumulative Distribution Function}
\newacronym{cn}{CN}{Core Network}
\newacronym{cqi}{CQI}{Channel Quality Information}
\newacronym{csirs}{CSI-RS}{Channel State Information - Reference Signal}
\newacronym{dc}{DC}{Dual Connectivity}
\newacronym{dce}{DCE}{Direct Code Execution}
\newacronym{dci}{DCI}{Downlink Control Information}
\newacronym{dl}{DL}{Downlink}
\newacronym{dmr}{DMR}{Deadline Miss Ratio}
\newacronym{dmrs}{DMRS}{Demodulation Reference Signal}
\newacronym{e2e}{E2E}{End-to-End}
\newacronym{ecn}{ECN}{Explicit Congestion Notification}
\newacronym{edf}{EDF}{Earliest Deadline First}
\newacronym{enb}{eNB}{evolved Node Base}
\newacronym{epc}{EPC}{Evolved Packet Core}
\newacronym{es}{ES}{Edge Server}
\newacronym{fdma}{FDMA}{Frequency Division Multiple Access}
\newacronym{fdd}{FDD}{Frequency Division Duplexing}
\newacronym[firstplural=Radio Access Technologies (RATs)]{rat}{RAT}{Radio Access Technology}
\newacronym{fs}{FS}{Fast Switching}
\newacronym{ftp}{FTP}{File Transfer Protocol}
\newacronym{gnb}{gNB}{Next Generation Node Base}
\newacronym{harq}{HARQ}{Hybrid Automatic Repeat reQuest}
\newacronym{hetnet}{HetNet}{Heterogeneous Network}
\newacronym{hh}{HH}{Hard Handover}
\newacronym{hol}{HOL}{Head-of-Line}
\newacronym{iot}{IoT}{Internet of Things}
\newacronym{los}{LOS}{Line of Sight}
\newacronym{lte}{LTE}{Long Term Evolution}
\newacronym{m2m}{M2M}{Machine to Machine}
\newacronym{mac}{MAC}{Medium Access Control}
\newacronym{mcs}{MCS}{Modulation and Coding Scheme}
\newacronym{mec}{MEC}{Mobile Edge Cloud}
\newacronym{mi}{MI}{Mutual Information}
\newacronym{mimo}{MIMO}{Multiple Input, Multiple Output}
\newacronym{mptcp}{MPTCP}{Multipath TCP}
\newacronym{mr}{MR}{Maximum Rate}
\newacronym{mss}{MSS}{Maximum Segment Size}
\newacronym{mtd}{MTD}{Machine-Type Device}
\newacronym{mtu}{MTU}{Maximum Transmission Unit}
\newacronym{nfv}{NFV}{Network Function Virtualization}
\newacronym{nlos}{NLOS}{Non Line of Sight}
\newacronym{nr}{NR}{New Radio}
\newacronym{ofdm}{OFDM}{Orthogonal Frequency Division Multiplexing}
\newacronym{pdcch}{PDCCH}{Physical Downlonk Control Channel}
\newacronym{pdcp}{PDCP}{Packet Data Convergence Protocol}
\newacronym{pdsch}{PDSCH}{Physical Downlink Shared Channel}
\newacronym{pdu}{PDU}{Packet Data Unit}
\newacronym{pf}{PF}{Proportional Fair}
\newacronym{pgw}{PGW}{Packet Gateway}
\newacronym{phy}{PHY}{Physical}
\newacronym{pbch}{PBCH}{Physical Broadcast Channel}
\newacronym[plural=\gls{mme}s,firstplural=Mobility Management Entities (MMEs)]{mme}{MME}{Mobility Management Entity}
\newacronym{prb}{PRB}{Physical Resource Block}
\newacronym{pss}{PSS}{Primary Synchronization Signal}
\newacronym{pucch}{PUCCH}{Physical Uplink Control Channel}
\newacronym{pusch}{PUSCH}{Physical Uplink Shared Channel}
\newacronym{ran}{RAN}{Radio Access Network}
\newacronym{red}{RED}{Random Early Detection}
\newacronym{rlc}{RLC}{Radio Link Control}
\newacronym{rrc}{RRC}{Radio Resource Control}
\newacronym{rrm}{RRM}{Radio Resource Management}
\newacronym{rr}{RR}{Round Robin}
\newacronym{rs}{RS}{Remote Server}
\newacronym{rsrp}{RSRP}{Reference Signal Received Power}
\newacronym{rss}{RSS}{Received Signal Strength}
\newacronym{rtt}{RTT}{Round Trip Time}
\newacronym{rw}{RW}{Receive Window}
\newacronym{rx}{RX}{Receiver}
\newacronym{sack}{SACK}{Selective Acknowledgment}
\newacronym{sap}{SAP}{Service Access Point}
\newacronym{sch}{SCH}{Secondary Cell Handover}
\newacronym{scoot}{SCOOT}{Split Cycle Offset Optimization Technique}
\newacronym{sdma}{SDMA}{Spatial Division Multiple Access}
\newacronym{sinr}{SINR}{Signal to Interference plus Noise Ratio}
\newacronym{sm}{SM}{Saturation Mode}
\newacronym{snr}{SNR}{Signal to Noise Ratio}
\newacronym{son}{SON}{Self-Organizing Network}
\newacronym{ss}{SS}{Synchronization Signal}
\newacronym{sss}{SSS}{Secondary Synchronization Signal}
\newacronym{tb}{TB}{Transport Block}
\newacronym{tcp}{TCP}{Transmission Control Protocol}
\newacronym{tdd}{TDD}{Time Division Duplexing}
\newacronym{tdma}{TDMA}{Time Division Multiple Access}
\newacronym{tfl}{TfL}{Transport for London}
\newacronym{tm}{TM}{Transparent Mode}
\newacronym{trp}{TRP}{Transmitter Receiver Pair}
\newacronym{tti}{TTI}{Transmission Time Interval}
\newacronym{ttt}{TTT}{Time-to-Trigger}
\newacronym{tx}{TX}{Transmitter}
\newacronym{ue}{UE}{User Equipment}
\newacronym{ul}{UL}{Uplink}
\newacronym{uml}{UML}{Unified Modeling Language}
\newacronym{um}{UM}{Unacknowledged Mode}
\newacronym{utc}{UTC}{Urban Traffic Control}
\newacronym{vm}{VM}{Virtual Machine}
\newacronym{fw}{FW}{Flow Window}

\usepackage{placeins}
\newif\iftikz
\tikztrue
\begin{document}
\title{milliProxy: a TCP Proxy Architecture\\for 5G mmWave Cellular Systems}

\author{Michele Polese$^*$, Marco Mezzavilla$^\dagger$, Menglei Zhang$^\dagger$,\\Jing Zhu$^\diamond$, Sundeep Rangan$^\dagger$, Shivendra Panwar$^\dagger$, Michele Zorzi$^*$\\
 \small $^*$Department of Information Engineering, University of Padova, Italy - 
e-mail: \{polesemi, zorzi\}@dei.unipd.it\\
$^\dagger$NYU Wireless, New York University, NY, USA - 
e-mail: \{menglei, mezzavilla, srangan\}@nyu.edu, panwar@catt.poly.edu\\
$^\diamond$Intel Corporation - e-mail: jing.z.zhu@intel.com}



\maketitle

\tikzstyle{startstop} = [rectangle, rounded corners, minimum width=2cm, minimum height=0.5cm,text centered, draw=black]
\tikzstyle{io} = [trapezium, trapezium left angle=70, trapezium right angle=110, minimum width=3cm, minimum height=1cm, text centered, draw=black]
\tikzstyle{process} = [rectangle, minimum width=2cm, minimum height=0.5cm, text centered, draw=black, alignb=center]
\tikzstyle{decision} = [ellipse, minimum width=2cm, minimum height=1cm, text centered, draw=black]
\tikzstyle{arrow} = [thick,<->,>=stealth]
\tikzstyle{line} = [thick,>=stealth]
\tikzstyle{darrow} = [thick,<->,>=stealth,dashed]
\tikzstyle{sarrow} = [thick,->,>=stealth]
\tikzstyle{larrow} = [line width=0.05mm,dashdotted,>=stealth]

\begin{abstract}
TCP is the most widely used transport protocol in the internet. However, it offers suboptimal performance when operating over high bandwidth mmWave links. The main issues introduced by communications at such high frequencies are (i) the sensitivity to blockage and (ii) the high bandwidth fluctuations due to \gls{los} to \gls{nlos} transitions and vice versa. In particular, TCP has an abstract view of the end-to-end connection, which does not properly capture the dynamics of the wireless mmWave link. The consequence is a suboptimal utilization of the available resources. In this paper we propose a TCP proxy architecture that improves the performance of TCP flows without any modification at the remote sender side. The proxy is installed in the Radio Access Network, and exploits information available at the \gls{gnb} in order to maximize throughput and minimize latency.
\end{abstract}

\begin{picture}(0,0)(0,-345)
\put(0,0){
\put(0,0){\footnotesize This paper was presented at the 2017 51st Asilomar Conference} 
\put(0,-10){\footnotesize on Signals, Systems and Computers, Pacific Grove, CA, 2017.}
\put(0, -20){\footnotesize Copyright (c) 2017 IEEE.}}
\end{picture}

\section{Introduction}
Communication at mmWave frequencies represents the new frontier for wireless networks~\cite{rappaport1}. Nowadays, indeed, most wireless standards are constrained in a very small bandwidth which can be allocated in the spectrum below 6 GHz. The mmWave spectrum (and in general frequencies above 6 GHz) would allow wireless communications to benefit from wide chunks of untapped spectrum that could be used to reach very high data rates. 

Millimeter waves are being considered for the next generation of cellular networks (i.e., 5G), which is currently being standardized by 3GPP as \gls{nr}~\cite{38802}, as well as for public safety~\cite{access-critical} and vehicular communications~\cite{choi2016vehicular,giordani2017mocast}. Several applications in these scenarios require high data rate, and with low latency: consider for example real-time video streaming for remote interaction with autonomous robots, virtual reality applications or raw sensor data (e.g., from RADAR or LIDAR) exchanges between vehicles. However, the mmWave technology offers both a great potential and a number of challenges, which are related to the harsh propagation conditions that must be addressed before mmWaves can be reliably deployed. The very high path loss can be overcome with beamforming techniques, which in turn require the design of protocols that account for directionality~\cite{zorzimac}. Moreover, mmWaves are affected by blockage from obstacles (e.g., buildings, trees, the human body itself)~\cite{singh2007millimeter}, and this has an impact on the service availability and the achievable data rate. 

These limitations have an impact not only on the protocol stack of the wireless link but also on higher layers protocols, such as TCP, which is the most widely used transport protocol in the internet, for a variety of different applications that require flow control and reliable transmissions. TCP however suffers in terms of both reduced throughput and high latency when deployed over an end-to-end connection whose last link is based on mmWave radios~\cite{mmNet}, and it is not capable of delivering the high data rate and low latency performance that the aforementioned applications require~\cite{polese2017tcp}. 

In this paper we propose the design of \emph{milliProxy}, a novel TCP proxy for mmWave mobile networks aimed at fully reaping the benefits of mmWave links to achieve high throughput with low latency. It is transparent to the end hosts of the connection, and respects the end-to-end connection semantics. The main rationale is to split the TCP control loop in the mmWave RAN to optimize the flow control over the wireless link. It is based on a cross-layer, data driven approach and enables a number of optimizations for the operation of TCP in mmWave networks. 

The rest of the paper is organized as follows. In Sec.~\ref{sec:soa}, we extensively discuss the main limitations of TCP over mmWave links, and provide an overview of the literature related to TCP proxies for traditional wireless networks. The architecture of milliProxy is described in Sec.~\ref{sec:architecture}, and the results of a performance evaluation campaign are reported in Sec.~\ref{sec:results}. Finally, conclusions and future extensions of this work are provided in Sec.~\ref{sec:concl}.

\section{Related Work}\label{sec:soa}

\subsection{5G mmWave networks and challenges for TCP}
In this section, we provide an overview of the main challenges related to the usage of TCP in mmWave cellular networks.
They can potentially enable gigabit-per-second cell data rates thanks to the very large bandwidth available~\cite{rappaport1}, but the deployment of a reliable 5G mmWave network is challenging, mainly because of the harsh propagation environment at these frequencies. The high propagation loss can be compensated using beamforming techniques, with a large number of antennas in the transceivers, but this requires the design of \gls{phy} and \gls{mac} layer protocols able to cope with directionality~\cite{zorzimac}. Moreover, mmWave frequencies are sensitive to blockage from a wide range of materials~\cite{singh2007millimeter}, and, while it is possible to transmit and receive data also in \gls{nlos} exploiting reflections, there is a massive difference in the achivable data rate and reliability when transitioning from \gls{los} to \gls{nlos} (with differences in the \gls{sinr} in the order of 30 dB) and vice versa.

Therefore, it is challenging to completely benefit from the vast amount of resources, especially when taking into account the end-to-end performance and the complex interaction between higher-layer protocols and the mmWave stack, as shown in previous works~\cite{menglei2016bufferbloat,mmNet,polese2017mptcp,polese2017mobility,saha2015wifi,sur2017wifi}. In particular, the main issues related to the most widely used transport protocol in end-to-end networks (i.e., TCP) are related to the slow reactiveness of TCP with respect to changes in the channel quality of the mmWave link~\cite{menglei2016bufferbloat,mmNet,polese2017mobility}. This translates into a sub-optimal utilization of the available resources, since the ramp up of the congestion window of the most used congestion control algorithms (e.g., TCP CUBIC~\cite{ha2008cubic} and TCP NewReno~\cite{NewReno}) is too slow and limits the achievable goodput. This issue is more marked in mmWave links than in traditional sub-6 GHz deployments, because of an orders-of-magnitude difference in bandwidth (and thus achievable rate), and it becomes more relevant as the \gls{rtt} of the connection increases, as shown in~\cite{polese2017mobility}. Another consequence is the emergence of the bufferbloat phenomenon~\cite{mmNet}. When the channel condition changes from \gls{los} to \gls{nlos}, there is a drop in the data rate offered by the physical layer. This is a problem because the packets in excess are buffered at the \gls{rlc} layer, and the TCP sender is not aware of the update until an \gls{aqm} mechanism or a buffer overflow drops one or more packets. This causes an increase in the buffer occupancy and, consequently, in the end-to-end latency. Finally, there is the possibility of extended outages, due to blockage and the lack of a fall-back link. TCP reacts to these events with retransmission timeouts, and halves the \textit{slow start threshold} at each of these events. Then, when the connection resumes, the duration of the slow start phase (with the exponential increase of the congestion window) is limited, and the TCP sender stays in congestion avoidance (with a linear increase of the congestion window) for most of the time, thus exacerbating the issue related to the slow congestion window ramp up.

\subsection{TCP Performance Enhancing Proxies}
The performance of TCP on wireless networks has been under the spotlight since the 1990s, when the first cellular networks capable of data transmission were commercially deployed. Even though TCP faces more challenging conditions when running on top of mmWave cellular networks, it is worth describing the main approaches that can be found in the literature related to the enhancement of TCP performance on wireless links. 

A first comprehensive review on the topic can be found in a paper by Balakrishnan \textit{et al.}~\cite{balakrishnan1997comparison}. The authors claim that the poor performance of TCP in mobile networks is due to packet losses over an unreliable channel. However, as shown in~\cite{polese2017mptcp}, the channel losses can be masked by retransmission mechanisms. Moreover, the considered links have very low data rate and small buffers are used in the network. The settings in a mmWave networks are very different, since large buffers and retransmissions are already implemented in the wireless link to make up for packet loss at the price of increased latency and exposing more the network to the bufferbloat phenomenon. However, the authors of~\cite{balakrishnan1997comparison} provide a comparison of different strategies that can possibly be adapted to mmWave networks, using TCP Reno as a baseline, and including also TCP split approaches. 



In a more recent paper~\cite{liu2016improving}, Liu \textit{et al.} introduce a TCP proxy middlebox for the optimization of TCP performance without the need for any modification to the protocol stack of servers, clients and base stations. They observe that the adoption of a new end-to-end TCP congestion control mechanism may be useless in the presence of HTTP proxies, which are frequently used in mobile networks. Moreover, they design their solution for modern LTE networks, characterized by large buffers (in the order of 5 MB) and bandwidth fluctuations (even if not as wide as those in mmWave networks~\cite{polese2017tcp,rappaport1}), and a fixed network which does not act as a bottleneck. Their solution is a middlebox that can be placed anywhere in the mobile operator core network, and breaks the TCP connection in two segments, i.e., it does not respect end-to-end connection semantics\footnote{According to~\cite{rfc3135,saltzer1984end}, the end-to-end principle, which states that certain functions in the internet are designed to work at the end hosts, is a founding paradigm of the internet. A proxy that splits the TCP connection into two independent segments does not respect the end-to-end TCP semantics, meaning that ACKs may be sent to the sender before the packet is actually received by the other end host.}.
This box performs some optimizations on the fly, such as (i) not using the information of the receiver congestion window, which may be too small with respect to the actual rate available on the link, given that experimental evaluations on the receiver buffer in real devices have highlighted that it is never filled; (ii) changing retransmission patterns by intercepting duplicate ACKs; (iii) tuning the congestion window with a rate estimation algorithm. In this design, the TCP connection from the sender to the receiver is terminated at the middlebox, which buffers the packets for the final receiver until it can forward them. 

A third approach is described in~\cite{ren2011modeling}, where Ren \textit{et al.} introduce a TCP proxy in the mobile network base station. This study, however, is focused on the UMTS architecture.
Their approach is based on a queue control mechanism: by using the sliding mode variable structure (SMVS) control theory the buffer queue length at the base station is kept at the same size. This proxy does not respect the end-to-end TCP semantics, because it terminates the connection at the proxy. The advertised window at the proxy is used to limit the sending rate of the server and to avoid buffering delays. At the proxy, a control mechanism is used to keep the queue length at a reference value, by inferring the bandwidth available at the base station. 

Some other interesting approaches that respect the end-to-end TCP semantics are (i) Mobile TCP (M-TCP)~\cite{brown1997m}, which freezes the TCP sender when it senses imminent congestion, in order to avoid packet loss and connection timeouts; and (ii) Snoop~\cite{balakrishnan1995improving}, also from Balakrishnan \textit{et al.}, which performs local retransmissions when TCP packet losses are sensed, in order to improve the connection reactiveness. I-TCP~\cite{bakre1995tcp}, instead, is a TCP split approach, not compliant with the end-to-end TCP semantics, that uses a traditional TCP congestion control also on the wireless link and does not yield a great performance improvement.

In~\cite{kim2017enhancing}, a performance enhancement proxy for mmWave cellular networks is proposed. It is installed in the base stations, and breaks the end-to-end TCP semantics by sending early ACKs to the server. Moreover it performs batch retransmissions, i.e., it retransmits the packets that were detected as lost as well as the segments with a sequence number which is close to that of the lost packets. The detection of the lost packets however assumes that on the link only \gls{harq} retransmissions are performed, while in general with TCP the \gls{rlc} \gls{am} is also used.
In the performance evaluation, moreover, the authors of~\cite{kim2017enhancing} limit the application data rate to 100 Mbit/s, which can be usually sustained also in \gls{nlos}. Therefore, the performance analysis does not account for the very high data rates that can be achieved with mmWave and for the wide rate variations of the \gls{los} to \gls{nlos} (or vice versa) transitions. Finally, it focuses only on the throughput and delivery ratio for a single user, without considering the latency and thus the bufferbloat problem. 

\section{End-to-end Proxy Architecture for mmWave}\label{sec:architecture}
In this section we describe our TCP proxy architecture for mmWaves, called milliProxy, and highlight the main innovations with respect to the solutions reported in Sec.~\ref{sec:soa}. Importantly, by being transparent to both the end points of the TCP flow, milliProxy respects the end-to-end semantics of the TCP connection, as opposed to most of the proposed approaches cited in Sec.~\ref{sec:soa}. The key functionalities of milliProxy are (1) the ability to split the control loop of the connection with a different and tunable \gls{fw} policy at the source server and at the proxy, as well as (2) the capability of controlling the \gls{mss} of the connection in the portion between the proxy and the \gls{ue}. 

\subsection{Proxy Architecture}
MilliProxy is a TCP proxy which can be implemented and deployed as a network function, composed of several modules that can be updated or changed. 
It can be placed in the \gls{gnb}, fully benefiting from the interaction with the mmWave protocol stack, or in a node in the core network, sharing out-of-band information with the \gls{gnb} to which the TCP receiver is connected. According to the position of the proxy, there may be the need to design a mechanism to cope with the user mobility. For example, if the proxy is in the \gls{gnb}, when the \gls{ue} performs a handover the network has to transfer the milliProxy's state from the source to the target \gls{gnb}. If instead the proxy is in an edge node of the core network, then it can manage multiple cells without the need to forward the state for each \gls{ue} handover. Additional considerations on this issue are left for future work.

\begin{figure}[t]
	\centering
	\includegraphics[width=\columnwidth]{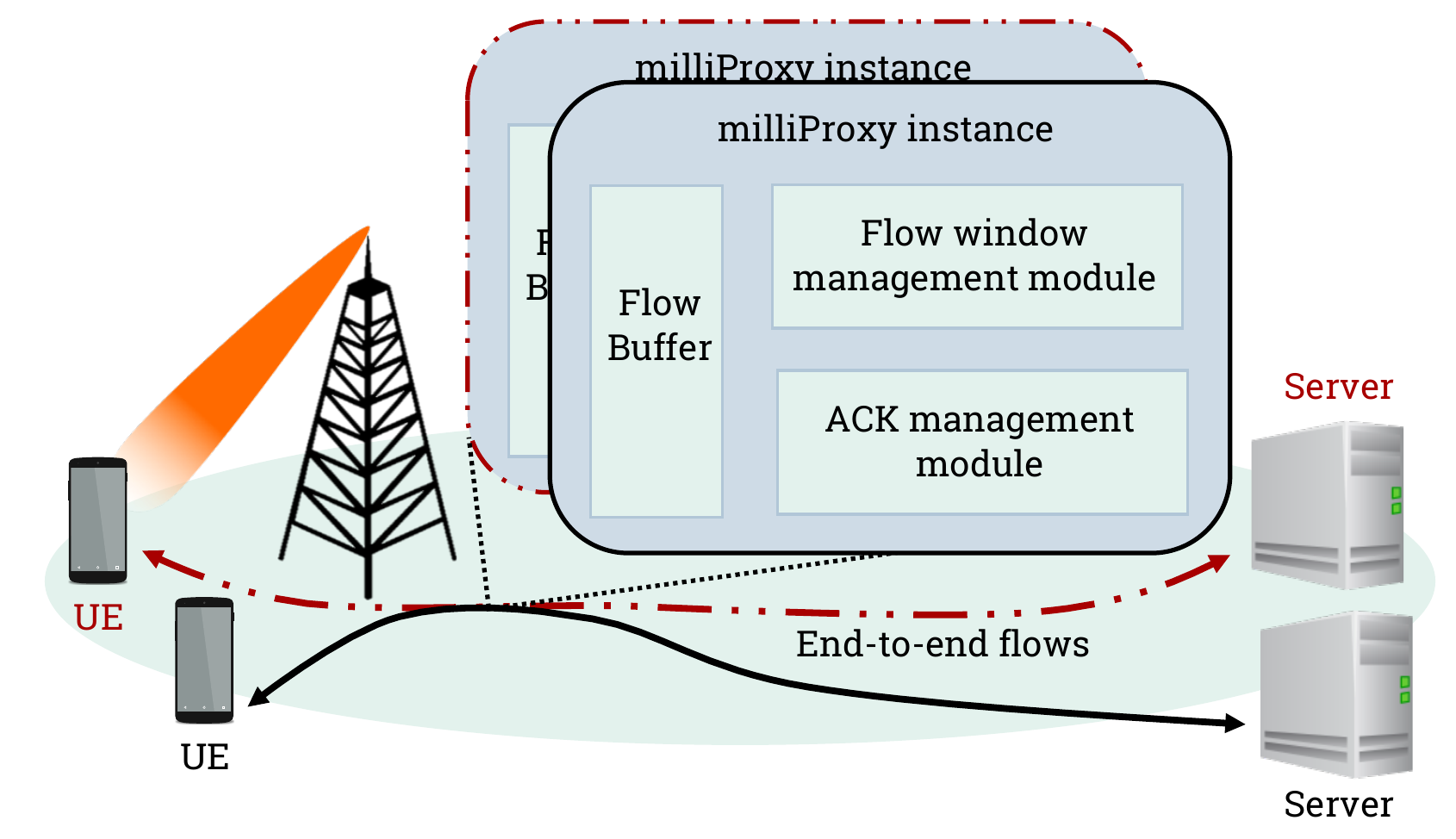}
	\caption{Architecture of milliProxy}
	\label{fig:mpArch}
\end{figure}

\begin{figure*}
	\centering
	\includegraphics[width=0.85\textwidth]{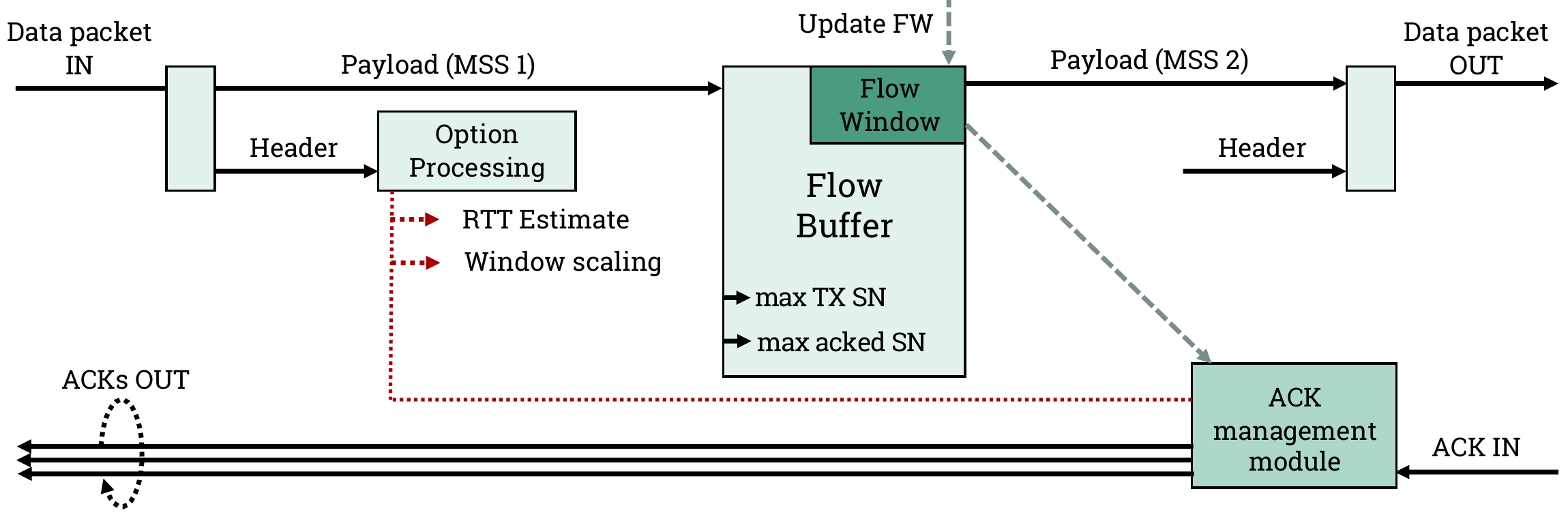}
	\caption{Packets processing in a milliProxy instance.}
	\label{fig:flow}
\end{figure*}

The basic structure of the proxy is shown in Fig.~\ref{fig:mpArch}. An instance of the proxy is created for each TCP flow which goes through the node in which it is installed, so that different policies can be enabled for different users, or different flows of the same user. Each instance has its own customizable buffer (set by default to 10 MB), flow window module and ACK management unit. The buffer is used to store the payload of the TCP packets before they can be delivered to the TCP receiver, and the ACK management unit checks for incoming ACKs to clear the contents of the buffer. The flow window policy is the equivalent at the proxy of the congestion window mechanism at the TCP sender, i.e., it controls the amount of data that can be forwarded by the proxy. The policy is not hard-coded into the proxy, but is loaded as a module, according to the implementation of the TCP congestion control mechanisms in the Linux kernel~\cite{casoni2015implementation}. The ACK management unit of the proxy modifies the \textit{advertised window} in each ACK that is relayed to the server in order to enforce the proxy flow window value also at the TCP sender. According to~\cite{RFC5681}, the TCP sender selects the minimum between its congestion window and the received value of the advertised window as the maximum number of bytes it can send. Similarly to~\cite{menglei2016bufferbloat}, the advertised window in the modified ACKs is set to be equal to the flow window determined at the proxy. This makes it possible to capture both components of the network, and adapt accordingly: the wired part is regulated by the classical TCP congestion control selected, while the wireless channel is used in a cross-layer fashion by the proxy, which selects the proper value of the advertised window.

The presence of the buffer makes it possible to tune the \gls{mss} of the connection between the proxy and the \gls{ue} differently from that of the other part of the connection, enabling further optimizations. If the \gls{mss} of the overall connection is limited by the \gls{mtu} of some intermediate networks using ethernet as link layer technology (i.e., the \gls{mss} is at most 1460 bytes), then the proxy buffers the 1460-byte payloads, and can send a larger segment which aggregates multiple payloads of the end-to-end connection. For example, fourteen 1460-bytes payloads received back-to-back in a small time interval can be combined into a single 20440-bytes segment which is sent from the proxy to the \gls{ue}. This increases the efficiency of the transmission process in the last mile of the connection, i.e., in the mmWave wireless link, because of the smaller overhead of the TCP/IP headers (in the previous example, just one TCP/IP header is used instead of fourteen), and because fewer uplink resources have to be scheduled for the transmission of ACKs from the \gls{ue} \cite{zhang2017will}. Notice that aggregation is generally performed also at the \gls{rlc} and \gls{mac} layers of very high-bandwidth connections in order to improve the transmission efficiency~\cite{38322,skordoulis2008frame}, and the larger \gls{mss} helps also this process, since fewer concatenation and segmentation operations are required at the transmitter and the receiver. 


Fig.~\ref{fig:flow} depicts how a packet is processed by milliProxy. By design, it is completely transparent to the \gls{ue}, i.e., the TCP receiver. It intercepts all the packets belonging to the flows it is handling, and the payload of data packets is stored in the proxy buffer. Any options in the packet header are processed, for example to estimate the \gls{rtt}, as will be described in the following sections, or to handle the advertised window scaling. The payload will then be sent as part of a larger segment as soon as the flow window allows it. When an ACK is received, the proxy checks its sequence number, and marks the corresponding bytes in the buffer as received, which will then be discarded, allowing the flow window to advance. Consequently, a number of ACKs corresponding to the number of original packets received (approximatively equal to the ratio between the \gls{mss} of the proxy-\gls{ue} connection and that of the server-proxy connection) is sent to the TCP sender. In each ACK the \textit{advertised window} value is overwritten with the value of the flow window in the proxy.


\subsection{RTT estimation}
The estimation of the \gls{rtt} can be performed using the TCP \textit{timestamp} option~\cite{RFC7323}. This option is symmetric, i.e., it is added both to data packets at the TCP sender and to ACKs at the receiver. It has a total length of 10 bytes, and contains two timestamps. The first ($TS_{\rm val}$) is that of the clock of the end host that transmits the packet, the second ($TS_{\rm echo}$) is the $TS_{\rm val}$ of a recently received packet from the other end host. Its usage is advised in~\cite{RFC7323} in order to improve the TCP performance, in terms of both throughput and security. 

\begin{figure}[t]
	\centering
	\includegraphics[width=\columnwidth]{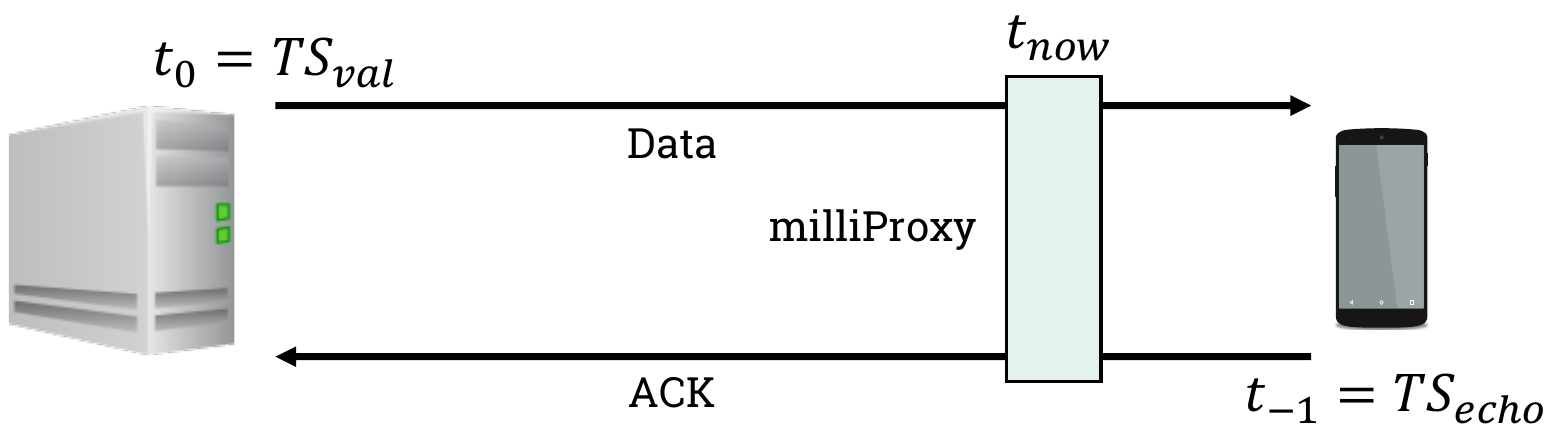}
	\caption{\gls{rtt} computation at the proxy.}
	\label{fig:ts}
\end{figure}

If both the end hosts share the same clock, the estimation of the \gls{rtt} is composed by two phases as follows. In the first one, which is shown in Fig.~\ref{fig:ts}, the milliProxy instance estimates the latency on the path from the UE to the server. The timestamp $TS_{\rm echo}$ of the data packet sent from the server to the \gls{ue} corresponds to the time $t_{-1}$ at which the \gls{ue} sent an ACK. Similarly, the timestamp $TS_{\rm val}$ in the same packet corresponds to the time instant $t_{0}$ at which the server transmitted the data packet. Given the very high packet rate that is sustained in mmWave networks, it is unlikely to observe a significant time interval between receiving the ACK corresponding to $TS_{\rm echo}$ and sending the data packet corresponding to $TS_{\rm val}$. Therefore, the latency of the uplink path can be estimated as $T_{\rm UE \rightarrow server} = t_0 - t_{-1}$. In a similar fashion, it is possible to use the timestamp values carried by ACK packets to estimate the latency on the downlink path $T_{\rm server \rightarrow UE}$. Finally, the \gls{rtt} is estimated as $RTT_e = T_{\rm server \rightarrow UE} + T_{\rm UE \rightarrow server}$. 

If instead the two end hosts do not have the same clock, or if the TCP timestamp option is not supported, other methods can be used to estimate the \gls{rtt} as reported in~\cite{veal2005new}. 

\subsection{Integration with the 5G protocol stack}
The proxy is configured to collect some statistics from the connected 5G \gls{gnb}. According to the location of the proxy, this data collection can be performed with or without delay. If the proxy is installed in the \gls{gnb}, the information can be retrieved instantaneously, whereas if it resides in a node in the core or edge network some signaling is necessary, which would introduce some incremental latency. Thanks to this information it is possible to enable a cross-layer approach, which is useful for the design of flow window management algorithms driven by the performance and the statistics of the mmWave link.

More information associated with each user can be retrieved from the protocol stack of the \gls{gnb}. The first is the \gls{rlc} buffer occupancy $B$, which can be seen as a signal of a congestion event and a consequent increase in latency. The second is an estimate of the \gls{phy} layer data rate between the \gls{ue} and the \gls{gnb}. In~\cite{liu2016improving} this is done by measuring the number of bytes transmitted in the previous slots, dividing it by the duration of the slots. This approach, however, is sensitive to the actual rate that is injected in the network by the TCP source, and can lead to an underestimation of the available rate if the source rate does not saturate the connection. This limitation is particularly relevant in mmWave networks, where it takes a long time for the TCP source to reach a full utilization of the available resources. In our previous works~\cite{menglei2016bufferbloat,azzino2017x}, instead, we rely on the information provided by the \gls{amc} module at the \gls{mac} layer. By knowing the channel quality of a \gls{ue} it is possible to compute the modulation and coding scheme, predict how many bytes the scheduler could allocate to the user (with full buffer assumption) in the next time slot, and divide by its duration to obtain an achievable data rate $R_{e}$ that is not influenced by the source rate. 
Another useful metric that can be acquired in a cross-layer setup is the \gls{sinr} of the \gls{ue}, which could give an indication on the link status: for example, if it is below a certain threshold, then the proxy will know that the \gls{ue} is in outage. 

\subsection{Window Management}\label{sec:window}
The management of the flow window is an essential component of milliProxy. In this paper we propose a scheme based on the computation of the \gls{bdp}. The implementation and testing of alternative \gls{fw} management policies is left for future works.

In the \gls{bdp}-based scheme, the \gls{fw} management module uses three different kinds of cross-layer data: the \gls{rlc} buffer occupancy $B$, the estimated data rate $R_{e}$ and the estimated \gls{rtt} $RTT_e$. The latter is filtered, and the minimum value $RTT_{\rm min}$ is selected following the approach in~\cite{bufferbloat3G,menglei2016bufferbloat,azzino2017x,cardwell2016bbr}, so that the queuing latency at the \gls{rlc} layer or at intermediate buffers is not taken into account. The flow window is then computed as $w = \lfloor RTT_{\rm min} R_e \rfloor $. When the \gls{rtt} estimate is not yet available (i.e., for the first ACK after the reception of the SYN packet), the flow window is arbitrarily initialized to a high value of 400 MB. 
Moreover, it is possible to make the policy more conservative when the \gls{rlc} buffer occupancy exceeds a predefined value (e.g., 2 MB). In this case, the flow window is set to $w = \max\{\lfloor RTT_{\rm min} R_e \rfloor - 2B, 0\}$.

\section{Performance Evaluation}\label{sec:results}

\subsection{ns-3 mmWave module}
In order to evaluate the performance of milliProxy in an end-to-end scenario, we implemented the proxy in ns-3~\cite{henderson2008network}, an open source network simulator which also features a mmWave cellular protocol stack. A complete description of the mmWave module is provided in~\cite{mezzavilla2017end,mmWaveSim}: it features a complete 3GPP-like protocol stack in the \glspl{gnb} and \glspl{ue}, with a custom PHY/MAC layers implementation, based on a dynamic \gls{tdd} scheme designed for low latency communications~\cite{dutta2017frame}, \gls{rlc}, \gls{pdcp} and \gls{rrc} layers. The channel model is based on the 3GPP channel model for frequencies above 6 GHz~\cite{38900}, and it models the time correlation among the channel impulse responses computed for moving users in order to account for spatial consistency. 

\begin{figure}[t]
\centering
\begin{tikzpicture}[font=\sffamily, scale=0.58, every node/.style={scale=0.58}]
  \centering

    \node[anchor=south west,inner sep=0] (image) at (0,0) {\includegraphics[width=0.7\textwidth]{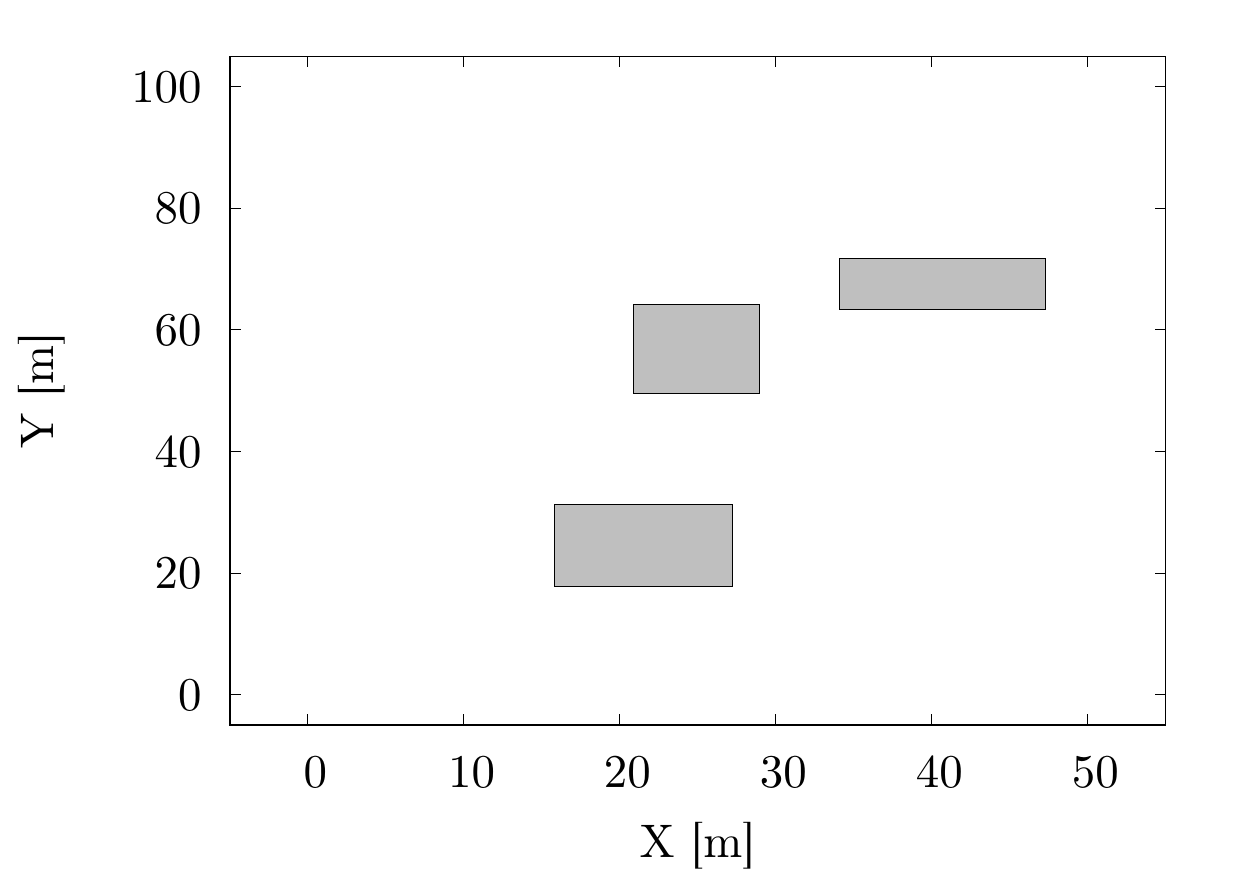}};
    \begin{scope}[x={(image.south east)},y={(image.north west)}]

        \filldraw[red,ultra thick] (0.56,0.905) circle (2pt);
        \node[anchor=south, align=left] at (0.572,0.77) (mm2label) {mmWave\\\gls{gnb}};

        \draw[sarrow] (0.248, 0.20) -- (0.88, 0.20);
        \node[anchor=south, text width=1.8cm] at (0.75, 0.20) (arrowLabel) {UE path at speed $v$};

        \filldraw[blue,ultra thick] (0.248, 0.20) circle (2pt);
        \node[anchor=south] at (0.248, 0.21) (ltelabel) {UE};

    \end{scope}   
\end{tikzpicture}
\caption{Randomly generated simulation scenario. The three grey rectangles represent obstacles such as buildings, cars, trees.}
\label{fig:map}
\end{figure}
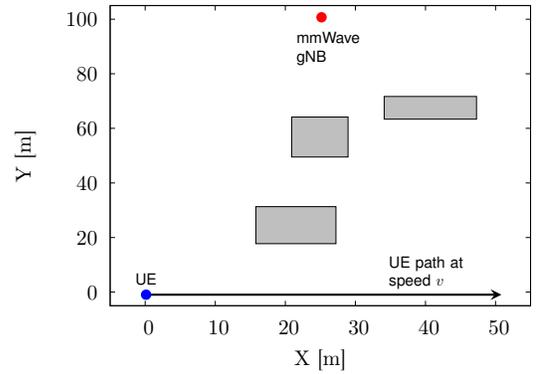

\begin{table}[t]
  \centering
  \small
  \begin{tabular}{@{}ll@{}}
    \toprule
    Parameter & Value \\
    \midrule
    mmWave carrier frequency & 28 GHz \\
    mmWave bandwidth & 1 GHz \\
    3GPP Channel Scenario & Urban Micro \\
    Max PHY layer rate & 3.2 Gbit/s \\
    S1 link latency $D_{S1}$  & 1 ms \\
    Latency from \acrshort{pgw} to server $D_{RS}$ & $[1, 5, 10, 20]$~ms\\
    RLC AM buffer size $B_{\rm RLC}$ & $[10, 20]$ MB \\
    RLC AM Reordering Timer & 1 ms \\
    RLC AM Report Buffer Status timer & 2 ms \\
    UE speed $v$ & 5 m/s \\
    TCP MSS$_1$ (server - proxy) & 1400 byte \\
    TCP MSS$_2$ (proxy - \gls{ue}) & 20000 byte \\
    \bottomrule
  \end{tabular}
  \caption{Simulation parameters}
  \label{table:params}
\end{table}

\subsection{Scenario and parameters}
The main simulation parameters are reported in Table~\ref{table:params}.
In this paper we focus on testing the performance of milliProxy in a single user scenario, in order to evaluate the responsiveness of the proxy architecture to channel variations, from \gls{los} to \gls{nlos} and viceversa. In order to model them, some obstacles are randomly deployed in the simulation scenario between the \gls{gnb} (which is at coordinates $(25, 100)$~m) and the \gls{ue} (moving from $(0, 0)$~m to $(50,0)$ at speed $v$). As the user moves, it will experience multiple transitions, with a random duration of each \gls{los} or \gls{nlos} phase in each different run of the simulation. An example of scenario is shown in Fig.~\ref{fig:map}. All the results are averaged over 50 independent simulation runs. 

\begin{figure}[t]
\centering
\begin{subfigure}[t]{\columnwidth}
	\centering
	\setlength\fwidth{0.85\columnwidth}
	\setlength\fheight{0.5\columnwidth}
%
%
\definecolor{mycolor1}{rgb}{0.24403,0.43583,0.99883}%
\definecolor{mycolor2}{rgb}{0.00357,0.72027,0.79170}%
\definecolor{lavender}{rgb}{0.9020,0.9020,0.9804}%
\definecolor{lightskyblue}{rgb}{0.6784,0.8471,0.9020}%
\definecolor{deepskyblue}{rgb}{0,0.7490,1}%
\definecolor{steelblue}{rgb}{0.2745,0.5098,0.7059}%
\definecolor{blue}{rgb}{0,0,1}%
\definecolor{royalblue}{rgb}{0.2549,0.4118,0.8824}%

\definecolor{gainsboro}{rgb}{0.8627,0.8627,0.8627}%
\definecolor{darkslategrey}{rgb}{0.1843,0.3098,0.3098}%
\definecolor{gray}{rgb}{0.5,0.5,0.5}%

\definecolor{lightcoral}{rgb}{0.9412,0.5020,0.5020}%
\definecolor{indianred}{rgb}{0.8039,0.3608,0.3608}%
\definecolor{lightsalmon}{rgb}{1.0000,0.6275,0.4784}%
\definecolor{darksalmon}{rgb}{0.9137,0.5882,0.4784}%
\begin{tikzpicture}
\pgfplotsset{every tick label/.append style={font=\scriptsize}}

\begin{axis}[%
width=0.951\fwidth,
height=\fheight,
at={(0\fwidth,0\fheight)},
scale only axis,
xtick=data,
xmin=1,
xmax=22,
ymin=250,
ymax=1900,
xlabel style={font=\footnotesize\color{white!15!black}},
xlabel={One-way end-to-end latency $D_{S1} + D_{RS}$ [ms]},
ylabel style={font=\footnotesize\color{white!15!black}},
ylabel={Goodput [Mbit/s]},
axis background/.style={fill=white},
xmajorgrids,
ymajorgrids,
ylabel shift = -5 pt,
yticklabel shift = -2 pt,
legend columns=2,
legend style={font=\scriptsize,at={(0.01,0.99)},anchor=south west,legend cell align=left,align=left,draw=white!15!black},
]
\addplot [color=steelblue, mark=x, line width=1pt, mark options={solid}]
 plot [error bars/.cd, y dir = both, y explicit]
 table[row sep=crcr, y error plus index=2, y error minus index=3]{%
2	1404.68925428412	176.855001957785	176.855001957785\\
6	1428.92697500849	164.612335980641	164.612335980641\\
11	1296.47412730566	130.174226110411	130.174226110411\\
21	1167.42564153603	143.598530918761	143.598530918761\\
};
\addlegendentry{milliProxy, $B=10$~MB}

\addplot [color=steelblue, dashed, mark=o, line width=1pt, mark options={solid}]
 plot [error bars/.cd, y dir = both, y explicit]
 table[row sep=crcr, y error plus index=2, y error minus index=3]{%
2	1176.3585636265	1415.83321856563	141.583321856563\\
6	846.794900411868	191.019117031338	191.019117031338\\
11	753.677388133778	189.381724669301	189.381724669301\\
21	520.464234969987	122.816181648	122.816181648\\
};
\addlegendentry{TCP NewReno, $B=10$~MB}

\addplot [color=indianred, mark=asterisk, line width=1pt, mark options={solid}]
 plot [error bars/.cd, y dir = both, y explicit]
 table[row sep=crcr, y error plus index=2, y error minus index=3]{%
2	1433.66605819281	171.013969156323	171.013969156323\\
6	1428.92697500849	166.318208858124	166.318208858124\\
11	1307.77899471757	91.8378398907634	91.8378398907634\\
21	1170.72302233603	145.459221406088	145.459221406088\\
};
\addlegendentry{milliProxy, $B=20$~MB}

\addplot [color=indianred, dashed, mark=diamond, line width=1pt, mark options={solid}]
 plot [error bars/.cd, y dir = both, y explicit]
 table[row sep=crcr, y error plus index=2, y error minus index=3]{%
2	1414.5602783331	157.893872248501	157.893872248501\\
6	1248.21785167668	184.877120007558	184.877120007558\\
11	1204.38450523967	186.333571009117	186.333571009117\\
21	588.268378514732	142.843928272843	142.843928272843\\
};
\addlegendentry{TCP NewReno, $B=20$~MB}

\addplot [color=darkslategrey, dotted, mark=v, line width=1pt, mark options={solid}]
 plot [error bars/.cd, y dir = both, y explicit]
 table[row sep=crcr, y error plus index=2, y error minus index=3]{%
2	1772.56983702558	164.424975953422	164.424975953422\\
6	1747.21131046073	175.569063818405	175.569063818405\\
11	1748.14333185386	178.187257511512	178.187257511512\\
21	1781.03197177138	166.011505676919	166.011505676919\\
};
\addlegendentry{UDP}

\end{axis}
\end{tikzpicture}%
	\caption{TCP goodput\vspace{0.2cm}}
	\label{fig:goodput}
\end{subfigure}
\begin{subfigure}[t]{\columnwidth}
	\centering
	\setlength\fwidth{0.85\columnwidth}
	\setlength\fheight{0.5\columnwidth}
%
%
\definecolor{mycolor1}{rgb}{0.24403,0.43583,0.99883}%
\definecolor{mycolor2}{rgb}{0.00357,0.72027,0.79170}%
\definecolor{lavender}{rgb}{0.9020,0.9020,0.9804}%
\definecolor{lightskyblue}{rgb}{0.6784,0.8471,0.9020}%
\definecolor{deepskyblue}{rgb}{0,0.7490,1}%
\definecolor{steelblue}{rgb}{0.2745,0.5098,0.7059}%
\definecolor{blue}{rgb}{0,0,1}%
\definecolor{royalblue}{rgb}{0.2549,0.4118,0.8824}%

\definecolor{gainsboro}{rgb}{0.8627,0.8627,0.8627}%
\definecolor{darkslategrey}{rgb}{0.1843,0.3098,0.3098}%
\definecolor{gray}{rgb}{0.5,0.5,0.5}%

\definecolor{lightcoral}{rgb}{0.9412,0.5020,0.5020}%
\definecolor{indianred}{rgb}{0.8039,0.3608,0.3608}%
\definecolor{lightsalmon}{rgb}{1.0000,0.6275,0.4784}%
\definecolor{darksalmon}{rgb}{0.9137,0.5882,0.4784}%
\begin{tikzpicture}
\pgfplotsset{every tick label/.append style={font=\scriptsize}}

\begin{axis}[%
width=0.951\fwidth,
height=\fheight,
at={(0\fwidth,0\fheight)},
scale only axis,
xtick=data,
xmin=1,
xmax=22,
ymin=0,
ymax=150,
xlabel style={font=\footnotesize\color{white!15!black}},
xlabel={One-way end-to-end latency $D_{S1} + D_{RS}$ [ms]},
ylabel style={font=\footnotesize\color{white!15!black}},
ylabel={RAN latency [ms]},
axis background/.style={fill=white},
xmajorgrids,
ymajorgrids,
yticklabel shift = -2 pt,
legend columns=2,
legend style={font=\scriptsize,at={(0.01,0.99)},anchor=south west,legend cell align=left,align=left,draw=white!15!black},
]
\addplot [color=steelblue, mark=x, line width=1pt, mark options={solid}]
 plot [error bars/.cd, y dir = both, y explicit]
 table[row sep=crcr, y error plus index=2, y error minus index=3]{%
2	3.79706078914194	2.42456483066931	2.42456483066931\\
6	10.0727212961421	4.44654237055533	4.44654237055533\\
11	17.5468107996325	7.23470845773903	7.23470845773903\\
21	20.8919542359608	5.40564170429527	5.40564170429527\\
};
\addlegendentry{milliProxy, $B=10$~MB}

\addplot [color=steelblue, dashed, mark=o, line width=1pt, mark options={solid}]
 plot [error bars/.cd, y dir = both, y explicit]
 table[row sep=crcr, y error plus index=2, y error minus index=3]{%
2	44.8082153011426	18.6110211861734	18.6110211861734\\
6	47.8926033178941	22.5976064770062	22.5976064770062\\
11	44.8741382003199	17.7365379420318	17.7365379420318\\
21	41.5503035710932	22.8446108341238	22.8446108341238\\
};
\addlegendentry{TCP NewReno, $B=10$~MB}

\addplot [color=indianred, mark=asterisk, line width=1pt, mark options={solid}]
 plot [error bars/.cd, y dir = both, y explicit]
 table[row sep=crcr, y error plus index=2, y error minus index=3]{%
2	2.67244552224907	1.03112118176732	1.03112118176732\\
6	10.0727212961421	4.49262176055559	4.49262176055559\\
11	18.6032800717946	8.29079523164865	8.29079523164865\\
21	20.8762573246491	5.46413578540927	5.46413578540927\\
};
\addlegendentry{milliProxy, $B=20$~MB}

\addplot [color=indianred, dashed, mark=diamond, line width=1pt, mark options={solid}]
 plot [error bars/.cd, y dir = both, y explicit]
 table[row sep=crcr, y error plus index=2, y error minus index=3]{%
2	115.796908541339	33.5470073596622	33.5470073596622\\
6	116.418185057814	28.3555632398154	28.3555632398154\\
11	101.9541580783	25.8667547838253	25.8667547838253\\
21	77.2169742316893	31.2171165342392	31.2171165342392\\
};
\addlegendentry{TCP NewReno, $B=20$~MB}

\end{axis}
\end{tikzpicture}%
	\caption{Latency in the RAN (from the PDCP at the eNB to that at the UE)}
	\label{fig:latency}
\end{subfigure}
\caption{Comparison of goodput and \gls{ran} latency with and without milliProxy, for different buffer sizes $B$.}
\label{fig:vsTCP}
\end{figure}
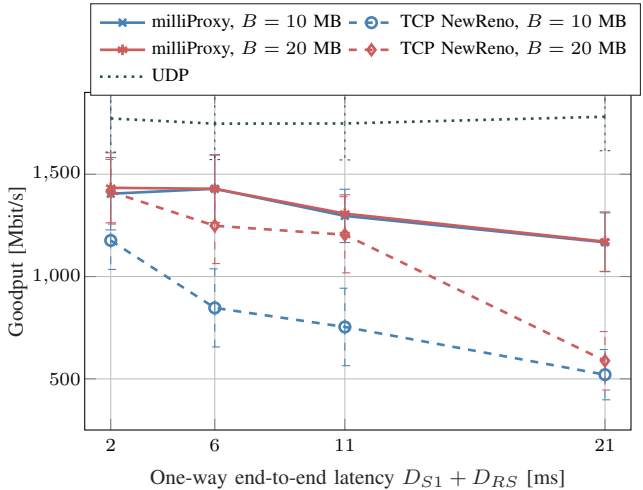
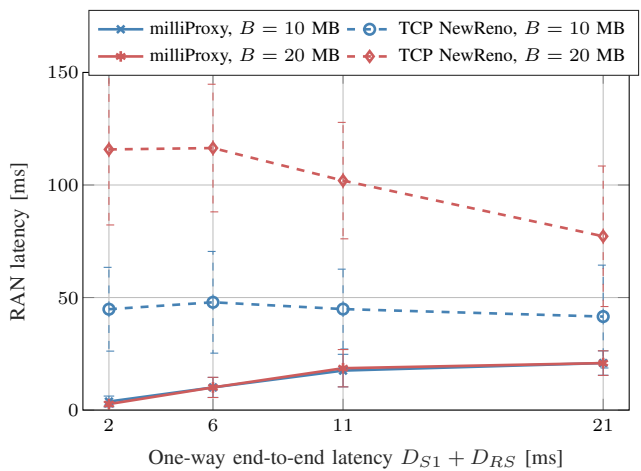

\begin{table}[t]

	\centering
	\begin{subfigure}[t]{\columnwidth}
	\centering
	  	\small
		\begin{tabular}{@{}lllll@{}}
			\toprule
			$D_{S1} + D_{RS}$ [ms] & 2 & 6 & 11 & 21 \\
			\midrule
			$B_{\rm RLC} = 10$~MB &  11.8008 & 4.7547 & 2.5574 & 1.9888 \\
			$B_{\rm RLC} = 20$~MB & 43.3299 &  11.5578  &  5.8104 &   3.6988 \\
			\bottomrule
		\end{tabular}
		\caption{\gls{ran} latency reduction when using milliProxy, i.e., ratio between the latency with TCP NewReno and that with milliProxy.	\vspace{0.2cm}}
		\label{table:latencyReduction}
	\end{subfigure}
	\begin{subfigure}[t]{\columnwidth}
	\centering
	\small
		\begin{tabular}{@{}lllll@{}}
			\toprule
			$D_{S1} + D_{RS}$ [ms] & 2 & 6 & 11 & 21 \\
			\midrule
			$B_{\rm RLC} = 10$~MB & 1.1941 & 1.6875 & 1.7202 & 2.2430 \\
			$B_{\rm RLC} = 20$~MB & 1.0135 & 1.1448 & 1.0765 & 1.9901 \\
			\bottomrule
		\end{tabular}
		\caption{TCP goodput gain when using milliProxy, i.e., ratio between the goodput with milliProxy and with TCP NewReno.}
		\label{table:latencyReduction}
	\end{subfigure}
	\caption{Goodput and latency performance gains with milliProxy.}
\end{table}

\subsection{Results}
Fig.~\ref{fig:vsTCP} shows a comparison of both goodput and \gls{ran} latency when milliProxy is deployed in the \gls{gnb} or not, for different \gls{rlc} buffer sizes $B_{\rm RLC}$ and fixed-network latencies. It can be seen that milliProxy performs better in terms of both goodput and latency, with a goodput gain of up to 2.24 times (combined with a latency reduction of 1.98 times) with the highest $D_{RS}$, or a latency reduction of 43 times with a similar goodput in the edge server scenario (i.e., $D_{RS} = 1$~ms). MilliProxy is therefore effective at reducing the impact of the bufferbloat issue: when the channel switches from a \gls{los} to a \gls{nlos} state, milliProxy can reduce the TCP sending rate faster, and thereby avoid extra queuing latency. On the other hand, when the channel quality improves, milliProxy is able to (i) track the available data rate at the physical layer and (ii) promptly inform the TCP sender of the increased resource availability, which indeed results in higher goodput. The performance of milliProxy is independent on the buffer size, since it manages to keep the buffer occupancy and consequently the \gls{rlc} queuing delay to a minimum. As shown in Fig.~\ref{fig:vsTCP} and extensively discussed in \cite{zhang2017will}, traditional approaches without proxy result in higher goodput at the price of increased \gls{ran} latency when using larger buffers.

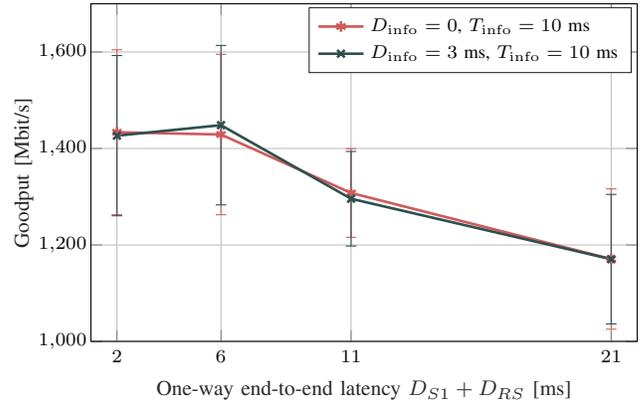
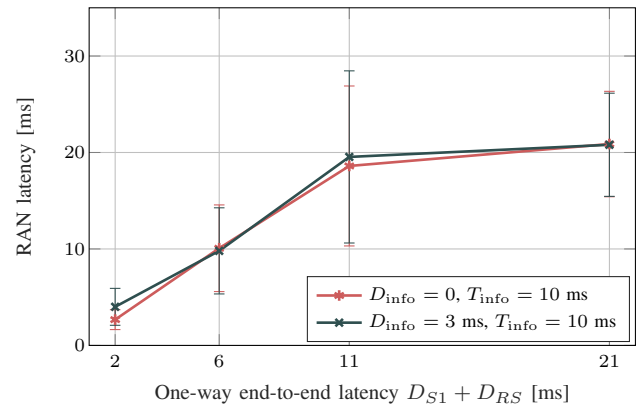
\begin{figure}[t]
\centering
\begin{subfigure}[t]{\columnwidth}
	\centering
	\setlength\fwidth{0.85\columnwidth}
	\setlength\fheight{0.5\columnwidth}
%
%
\definecolor{mycolor1}{rgb}{0.24403,0.43583,0.99883}%
\definecolor{mycolor2}{rgb}{0.00357,0.72027,0.79170}%
\definecolor{lavender}{rgb}{0.9020,0.9020,0.9804}%
\definecolor{lightskyblue}{rgb}{0.6784,0.8471,0.9020}%
\definecolor{deepskyblue}{rgb}{0,0.7490,1}%
\definecolor{steelblue}{rgb}{0.2745,0.5098,0.7059}%
\definecolor{blue}{rgb}{0,0,1}%
\definecolor{royalblue}{rgb}{0.2549,0.4118,0.8824}%

\definecolor{gainsboro}{rgb}{0.8627,0.8627,0.8627}%
\definecolor{darkslategrey}{rgb}{0.1843,0.3098,0.3098}%
\definecolor{gray}{rgb}{0.5,0.5,0.5}%

\definecolor{lightcoral}{rgb}{0.9412,0.5020,0.5020}%
\definecolor{indianred}{rgb}{0.8039,0.3608,0.3608}%
\definecolor{lightsalmon}{rgb}{1.0000,0.6275,0.4784}%
\definecolor{darksalmon}{rgb}{0.9137,0.5882,0.4784}%
\begin{tikzpicture}
\pgfplotsset{every tick label/.append style={font=\scriptsize}}

\begin{axis}[%
width=0.951\fwidth,
height=\fheight,
at={(0\fwidth,0\fheight)},
scale only axis,
xtick=data,
xmin=1,
xmax=22,
ymin=1000,
ymax=1700,
xlabel style={font=\footnotesize\color{white!15!black}},
xlabel={One-way end-to-end latency $D_{S1} + D_{RS}$ [ms]},
ylabel style={font=\footnotesize\color{white!15!black}},
ylabel={Goodput [Mbit/s]},
axis background/.style={fill=white},
xmajorgrids,
ymajorgrids,
ylabel shift = -5 pt,
yticklabel shift = -2 pt,
legend style={font=\scriptsize,at={(0.99,0.99)},anchor=north east,legend cell align=left,align=left,draw=white!15!black},
]
\addplot [color=indianred, mark=asterisk, line width=1pt, mark options={solid}]
 plot [error bars/.cd, y dir = both, y explicit]
 table[row sep=crcr, y error plus index=2, y error minus index=3]{%
2	1433.66605819281	171.013969156323	171.013969156323\\
6	1428.92697500849	166.318208858124	166.318208858124\\
11	1307.77899471757	91.8378398907634	91.8378398907634\\
21	1170.72302233603	145.459221406088	145.459221406088\\
};
\addlegendentry{$D_{\rm info} = 0$, $T_{\rm info} = 10$~ms}

\addplot [color=darkslategrey, mark=x, line width=1pt, mark options={solid}]
 plot [error bars/.cd, y dir = both, y explicit]
 table[row sep=crcr, y error plus index=2, y error minus index=3]{%
2	1426.53142259816	166.090585838796	166.090585838796\\
6	1448.32053551391	165.306733456108	165.306733456108\\
11	1295.77950212509	98.099845998313	98.099845998313\\
21	1170.48360867542	134.276689722768	134.276689722768\\
};
\addlegendentry{$D_{\rm info} = 3$~ms, $T_{\rm info} = 10$~ms}

\end{axis}
\end{tikzpicture}%
	\caption{TCP goodput\vspace{0.2cm}}
	\label{fig:goodput}
\end{subfigure}
\begin{subfigure}[t]{\columnwidth}
	\centering
	\setlength\fwidth{0.85\columnwidth}
	\setlength\fheight{0.5\columnwidth}
%
%
\definecolor{mycolor1}{rgb}{0.24403,0.43583,0.99883}%
\definecolor{mycolor2}{rgb}{0.00357,0.72027,0.79170}%
\definecolor{lavender}{rgb}{0.9020,0.9020,0.9804}%
\definecolor{lightskyblue}{rgb}{0.6784,0.8471,0.9020}%
\definecolor{deepskyblue}{rgb}{0,0.7490,1}%
\definecolor{steelblue}{rgb}{0.2745,0.5098,0.7059}%
\definecolor{blue}{rgb}{0,0,1}%
\definecolor{royalblue}{rgb}{0.2549,0.4118,0.8824}%

\definecolor{gainsboro}{rgb}{0.8627,0.8627,0.8627}%
\definecolor{darkslategrey}{rgb}{0.1843,0.3098,0.3098}%
\definecolor{gray}{rgb}{0.5,0.5,0.5}%

\definecolor{lightcoral}{rgb}{0.9412,0.5020,0.5020}%
\definecolor{indianred}{rgb}{0.8039,0.3608,0.3608}%
\definecolor{lightsalmon}{rgb}{1.0000,0.6275,0.4784}%
\definecolor{darksalmon}{rgb}{0.9137,0.5882,0.4784}%
\begin{tikzpicture}
\pgfplotsset{every tick label/.append style={font=\scriptsize}}

\begin{axis}[%
width=0.951\fwidth,
height=\fheight,
at={(0\fwidth,0\fheight)},
scale only axis,
xtick=data,
xmin=1,
xmax=22,
ymin=0,
ymax=35,
xlabel style={font=\footnotesize\color{white!15!black}},
xlabel={One-way end-to-end latency $D_{S1} + D_{RS}$ [ms]},
ylabel style={font=\footnotesize\color{white!15!black}},
ylabel={RAN latency [ms]},
axis background/.style={fill=white},
xmajorgrids,
ymajorgrids,
ylabel shift = +4 pt,
yticklabel shift = -2 pt,
legend style={font=\scriptsize,at={(0.99,0.01)},anchor=south east,legend cell align=left,align=left,draw=white!15!black},
]
\addplot [color=indianred, mark=asterisk, line width=1pt, mark options={solid}]
 plot [error bars/.cd, y dir = both, y explicit]
 table[row sep=crcr, y error plus index=2, y error minus index=3]{%
2	2.67244552224907	1.03112118176732	1.03112118176732\\
6	10.0727212961421	4.49262176055559	4.49262176055559\\
11	18.6032800717946	8.29079523164865	8.29079523164865\\
21	20.8762573246491	5.46413578540927	5.46413578540927\\
};
\addlegendentry{$D_{\rm info} = 0$, $T_{\rm info} = 10$~ms}

\addplot [color=darkslategrey, mark=x, line width=1pt, mark options={solid}]
 plot [error bars/.cd, y dir = both, y explicit]
 table[row sep=crcr, y error plus index=2, y error minus index=3]{%
2	4.00123033327808	1.91353597889515	1.91353597889515\\
6	9.80154193744112	4.46404943524468	4.46404943524468\\
11	19.5403325521095	8.92305887906242	8.92305887906242\\
21	20.7998763063115	5.3471208310537	5.3471208310537\\
};
\addlegendentry{$D_{\rm info} = 3$~ms, $T_{\rm info} = 10$~ms}

\end{axis}
\end{tikzpicture}%
	\caption{Latency in the RAN (from the PDCP at the eNB to that at the UE)}
	\label{fig:latency}
\end{subfigure}
\caption{Comparison of goodput and \gls{ran} latency with different milliProxy configurations. $D_{\rm info}$ represents the latency needed to forward the cross-layer information from the \gls{gnb} to milliProxy, $T_{\rm info}$ is the periodicity at which this information is collected.}
\label{fig:milliConf}
\end{figure}

A comparison between different configuration options for milliProxy is given in Fig.~\ref{fig:milliConf}. In particular, we are interested in studying the sensitivity of goodput and latency with respect to the delay $D_{\rm info}$ in the acquisition of the cross-layer information from the \gls{gnb}: it is equal to 0 when milliProxy is deployed in the \gls{gnb}, and greater than 0 when installed in a node in the core or edge network. We consider $D_{\rm info} = 3$~ms, i.e., we assume that the latency between the proxy deployed in the core/edge network and the \gls{gnb} will be smaller than 3 ms. As shown in Fig.~\ref{fig:milliConf}, the two tested configurations have a similar behavior in terms of both goodput and latency, showing that milliProxy is robust with respect to different possible deployments in the edge network or in the \glspl{gnb}.

  

\section{Conclusions}\label{sec:concl}
In this paper we introduced milliProxy, a novel proxy designed to enhance the performance of TCP in mmWave cellular networks. We described the main challenges related to the usage of TCP on top of mmWave links, and the main proxy designs from the literature. MilliProxy splits the TCP control loop in two segments, while keeping the end-to-end semantics of TCP. It has a modular design, which enables the use of different \gls{mss} values and flow window management algorithms in the two portions of the connection (i.e., wired and wireless). The window control policy can benefit from the interaction of milliProxy with the protocol stack of the mmWave networks, which enables cross-layer approaches. We showed how a \gls{fw} policy based on the \gls{bdp} of the end-to-end connection allows a reduction in latency of up to 10 times or an increase in goodput of up to 2 times with respect to traditional TCP NewReno, as well as a robustness with respect to where milliProxy is placed in the network. 

MilliProxy is an option to reach high goodput with low latency with TCP in mmWave networks. As part of our future work we will test the proxy performance in a wide variety of scenarios in ns-3, analyzing the performance with multiple users, and with different flow window policies, and will consider the implementation in a real setup. 

\bibliographystyle{IEEEtran}
\bibliography{main.bib}{}  

\end{document}